\documentclass[journal=jacsat,manuscript=article]{achemso}

\usepackage[version=3]{mhchem} 
\usepackage{color, soul}
\usepackage{graphicx}
\graphicspath{ {./Figures/} }

\author{Bogdan R. Borodin}
\affiliation{Ioffe Institute, Saint-Petersburg, 194021, Russia}
\email{brborodin@gmail.com}
\author{Fedor A. Benimetskiy}
\affiliation{ITMO University, Saint-Petersburg, 197101, Russia}
\author{Valery Yu. Davydov}
\affiliation{Ioffe Institute, Saint-Petersburg, 194021, Russia}
\author{Ilya A. Eliseyev}
\affiliation{Ioffe Institute, Saint-Petersburg, 194021, Russia}
\author{Alexander N. Smirnov}
\affiliation{Ioffe Institute, Saint-Petersburg, 194021, Russia}
\author{Dmitry A. Pidgayko}
\affiliation{ITMO University, Saint-Petersburg, 197101, Russia}
\author{Sergey I. Lepeshov}
\affiliation{ITMO University, Saint-Petersburg, 197101, Russia}
\author{Andrey A. Bogdanov}
\affiliation{ITMO University, Saint-Petersburg, 197101, Russia}
\author{Prokhor A. Alekseev}
\affiliation{Ioffe Institute, Saint-Petersburg, 194021, Russia}

\title{Strong photoluminescence enhancement in indirect bandgap MoSe$_2$ nanophotonic resonator}

\abbreviations{IR,NMR,UV}
\keywords{American Chemical Society, \LaTeX}

\begin{document}

\begin{abstract}
Transition metal dichalcogenides (TMDs) are promising for new generation nanophotonics due to their unique optical properties. However, in contrast to direct bandgap TMDs monolayers, bulk samples have an indirect bandgap that restricts their application as light emitters. On the other hand, the high refractive index of these materials allows for creating of high-Q resonators. In this work, we proposed a method for nanofabrication of microcavities from the indirect TMD multilayer flakes making possible the pronounced resonant photoluminescence enhancement due to the cavity modes. We demonstrate this effect by the examples of whispering gallery mode (WGM) resonators fabricated from the bulk MoSe$_2$ using scanning probe lithography. Micro-photoluminescence ($\mu$-PL) investigation revealed WGM spectra of resonators with an enhancement factor up to 100. The characteristic features of WGMs are clearly seen from the scattering experiments and confirmed by numerical simulations. We believe that the suggested approach and structures have great prospectives in nanophotonics.
\end{abstract}

\section{Introduction}
Since the graphene discovery in 2004\cite{novoselov2004electric}, layered materials have become one of the most booming topics in many fields, such as materials science\cite{manzeli20172d}, condensed matter physics\cite{kennes2021moire}, optoelectronics\cite{jing2020tunable}, photonics\cite{mak2016photonics}, etc. The most perspective semiconductive materials seem to be transition metal dichalcogenides (TMDCs). This is due to the unique properties of their monolayers such as extraordinary light absorption\cite{bernardi2013extraordinary}, large exciton binding energy\cite{ugeda2014giant,hanbicki2015measurement}, strong and tailoring photoluminescence\cite{tonndorf2013photoluminescence,tongay2013broad,benimetskiy2019measurement}, the possibility of creating van der Waals heterostructures\cite{geim2013van,fan2020transfer}, twisting engineering\cite{michl2021intrinsic,tran2020moire,shabani2021deep}, etc. Such an active investigation of TMDCs monolayers has revived interest in studying the properties of bulk samples. These materials in their bulk form were actively studied back in the 70s.\cite{goldberg1975low,beal1979kramers,wilson1969transition,anedda1980exciton} However, today, researchers can take a fresh look at their bulk properties using the progress in understanding physical phenomena and the best modern equipment that allows looking deeper at known properties and finding new ones. Most recent studies revealed many new outstanding properties of multilayered TMDCs, such as giant optical anisotropy\cite{ermolaev2021giant}, polarizing effect\cite{berahim2019polarizing}, anapole-exciton polaritons\cite{verre2019transition}, exciton-plasmon-polaritons\cite{zhang2020hybrid}, exciton-polariton transport\cite{hu2017imaging}, second harmonic generation\cite{busschaert2020transition}, etc. All the above-mentioned make these materials a perfect candidate for a variety of nanophotonic applications including lasers\cite{ye2015monolayer,li2017room}, waveguides\cite{hu2017imaging,fei2016nano,munkhbat2022nanostructured}, high harmonic generation\cite{munkhbat2022nanostructured,khan2022optical,autere2018nonlinear}, bound states in the continuum\cite{muhammad2021optical,bernhardt2020quasi}, etc. However, there is a significant obstacle in the way of the implementation of light‐emitting nanophotonic devices. Multilayered TMDCs have an indirect bandgap that results in negligible photoluminescence (PL)\cite{tongay2012thermally}. To use some of the unique properties of TMDCs, many researchers integrate direct bandgap TMDC monolayers into external photonic circuits/resonators as a source of excitonic photoluminescence\cite{wang20192d,krasnok2018nanophotonics,ye2015monolayer}. However, because of the thickness, monolayers can not accommodate either waveguide or resonant modes in the visible and near-IR ranges. That requires forming photonic circuits and exciton/PL sources using various technological processes and different materials. All this considerably complicates the on-chip integration of TMDCs materials. Another way to solve this problem is to use the Purcell effect to enhance PL intensity of multilayered TMDCs\cite{purcell1995spontaneous,eswaramoorthy2021engineering}. This approach was successfully used to enhance light emission of Si in optical cavities\cite{cho2013silicon,fujita2013nanocavity,valenta2019nearly,gong2010photoluminescence}. Although due to strong free charge carriers absorption in Si, the effect was not game-changing\cite{kekatpure2008quantification,fauchet1998integration,elliman2007waveguiding,ridley2013quantum}. Owing to an extremely high refractive index in the visible and near-IR ranges (n $\approx 4-5$)\cite{hsu2019thickness,jung2019measuring}, high-Q nanocavities with a strong Purcell effect and enhanced emissivity might be made from bulk TMDCs (the principle is illustrated in Figure 1).

\begin{figure}[t!]
    \includegraphics[width=0.95\textwidth]{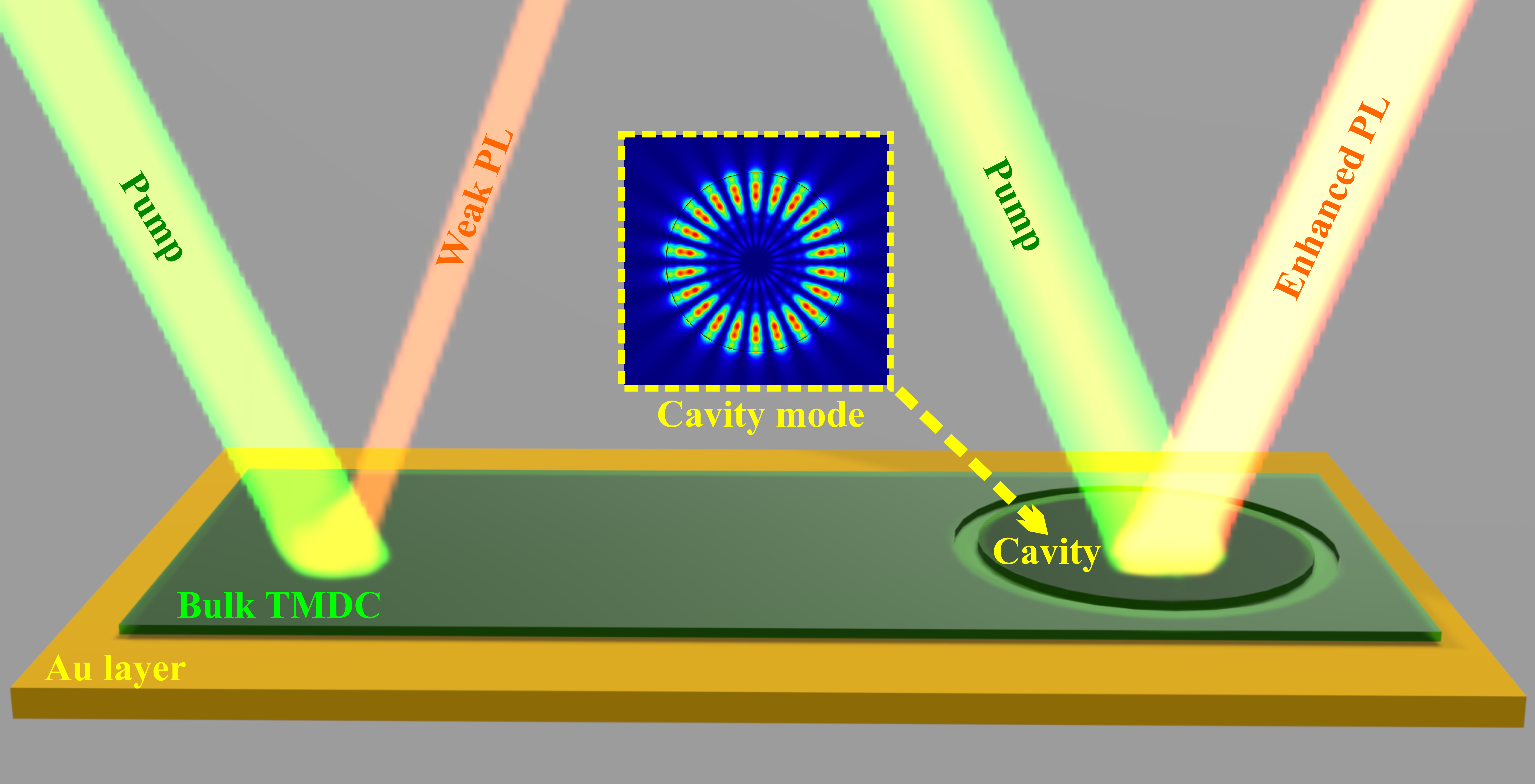}
    \caption{Principal scheme of PL enhancement in cavities made of indirect bandgap TMDCs. Owing to the indirect bandgap, the PL signal from bulk flakes is weak. However, it can be enhanced by a cavity.}
    \label{Principal scheme}
\end{figure}

In this work, we create disk whispering-gallery mode (WGM) optical nanoresonators from multilayered MoSe$_2$. Resistless mechanical probe lithography was used to fabricate nanoresonators so as not to disturb the pristine properties of TMDCs, which are highly sensitive to any contamination or chemical treatment\cite{du2008approaching,schwartz2019chemical,garcia2014advanced}. The obtained nanocavities demonstrate strongly enhanced (by two orders of magnitude) photoluminescence in the range from 850 to 1050 nm. The spectral features correspond to WGM resonances, which were confirmed by numerical simulations and scattering experiments. The results of the work allow introducing a novel type of stand alone TMDCs nano- and microcavities as a source of excitonic photoluminescence for on-chip integrated nanophotonic circuits.

\section{Results and Discussion}

\subsection{Fabrication of nanocavities}

\begin{figure}[b!]
    \includegraphics[width=\textwidth]{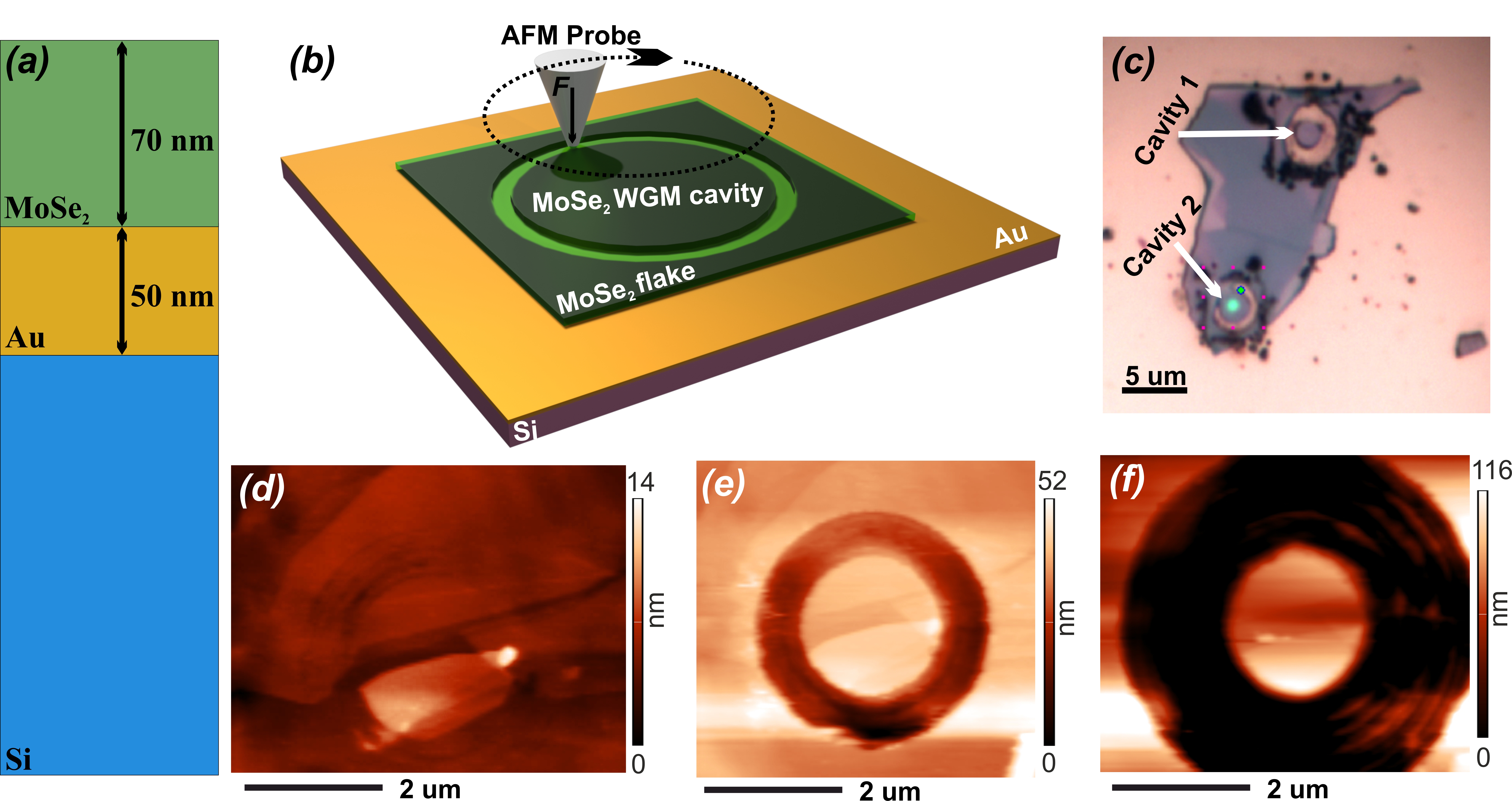}
    \caption{(a) Cross-section of the sample. (b) Scheme of frictional probe lithography of a cavity. (c) Optical image of the MoSe$_2$ flake with two cavities. (d-f) AFM images of the cavity creation process.}
    \label{fabrication}
\end{figure}

An experimental structure consisted of thin MoSe$_2$ flakes transferred on a Si substrate covered with 50 nm of gold. MoSe$_2$ flakes were obtained by micro-mechanical exfoliation and transferred using a standard approach (i.e., scotch-tape method) without using PDMS to prevent contamination with the polymer\cite{schwartz2019chemical}. This process is detailed in \textbf{Methods}. Figure \ref{fabrication}a shows the cross-section of the structure used.

To create a cavity, we firstly found a flake of a thickness about 70-100 nm that is sufficient to accommodate the WGMs. Then we used a frictional scanning probe lithography (f-SPL) to fabricate the cavities\cite{borodin2021mechanical2}. The scheme of f-SPL is shown in Figure \ref{fabrication}b. f-SPL is a resistless method of lithography based on the mechanical influence of an  atomic force microscopy (AFM) probe on the sample surface to remove the material (mechanical-SPL). However, while m-SPL conventionally uses high pressure to deepen lithographic patterns ("cutting" regime), f-SPL consists in consequent repetitions of lithographic patterns with small pressure on the sample. Thus, during f-SPL, the material is gradually rubbed out from the surface. This approach allows avoiding the cantilever twisting that prevents the formation of artifacts and makes it possible to maintain high resolution even in the case of thick samples (detailed in \textbf{Methods}). Figure \ref{fabrication}c shows an optical image of the processed flake. It can be seen that two circular cavities are formed from the flake, and the removed material is nearby. Figure \ref{fabrication}d-f demonstrates AFM images of the cavity creation process. The first is a relatively flat surface of the flake. The second is the surface with a half-thickness trench. The third is the fully separated cavity and the full-thickness trench. The optical properties of such cavities were studied by $\mu$-PL and dark-field spectroscopy.

\subsection{Optical properties of nanocavities}

\begin{figure}[t!]
    \centering
    \includegraphics[width=\textwidth]{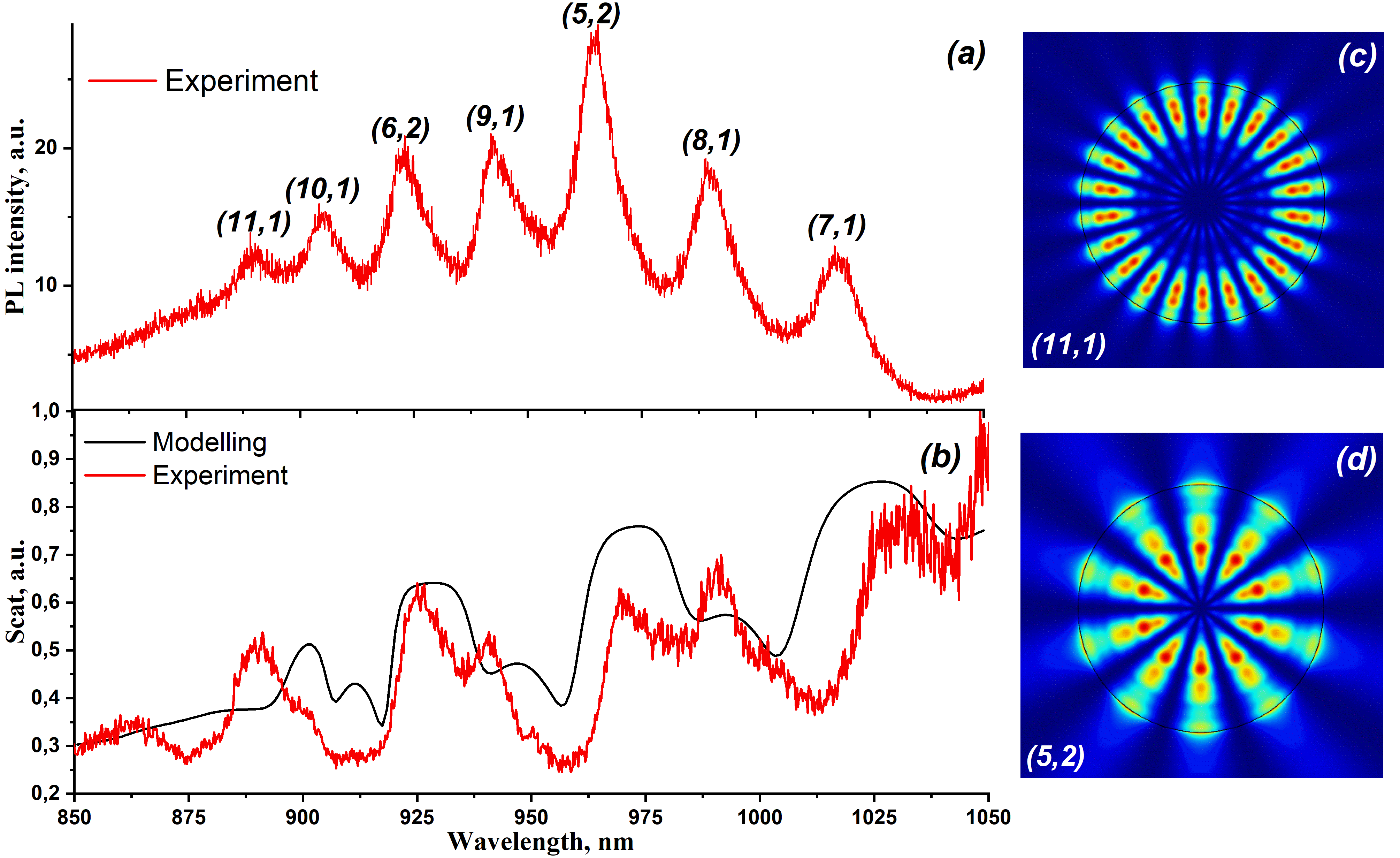}
    \caption{(a) Photoluminescence and (b) dark-field spectra of the cavity for s-polarised incidence. The red curves correspond to the experimental measurements, the black curve represents the numerical calculations.Numbers in brackets represent azimuthal and radial numbers (M, n) for modes corresponding to associated peaks. (c-d) Electrical field distribution in the cavity for (11, 1) and (5, 2) modes modeled in COMSOL Multiphysics. }
    \label{PL}
\end{figure}

As discussed in the Introduction, TMDs have many peculiar properties that can appear in the optical response of such structures. However, a combination of photoluminescence and scattering investigations is a reliable instrument to determine the nature of observed features. The shape of spectra and peak-to-peak distance are strong evidence of resonant phenomena, whether WGM or Mie resonances. Thus, the optical properties of the cavities were studied using micro-PL and dark-field spectroscopy. Figure~\ref{PL} demonstrates the results of the PL and experimental dark-field (DF) spectroscopy measurements supported by DF numerical calculations. In the $\mu$-PL spectrum, we observe a series of peaks (see Figure \ref{PL}a), which potentially may originate from the whispering gallery mode resonance. Further, we carry out the DF spectroscopy of the cavity previously exposed by PL measurements. Since the DF spectroscopy deals only with optical resonances of a cavity, it allows to unambiguously identify whether PL peaks have the resonant nature or not. The results are shown in Figure~\ref{PL}(b, red curve). One can see that we observe the resonant features in the spectral regions similar to that we get through PL measurements (Figure~\ref{PL}a). 
To prove the optical resonance nature of the PL peaks, we prepare comprehensive numerical simulations of optical scattering in COMSOL Multiphysics. We assume a MoSe$_2$ disk on the top of a Si substrate covered by a layer of Au. The permittivities of MoSe$_2$, Si and Au are taken from refs.~\cite{schinke2015uncertainty,beal1979kramers,johnson1972optical}. The disk diameter is 2.2~$\mu$m, the height is 70~nm, and the thickness of the Au layer is 50~nm. To calculate the scattering spectrum, we illuminate the structure by the s-polarised plane wave with the incident angle of 65$^0$ and collect the scattered wave in the numerical aperture of 0.65. The results of the numerical calculation of the DF spectrum are presented in Figure~\ref{PL}(b, black curve). The experimentally obtained DF spectrum is in a good agreement with the numerical one. Based on the resonant behaviour of the DF spectra and peak-to-peak comparison, we conclude that the PL peaks have the optical resonance nature. Figures~\ref{PL}c-d demonstrate electrical field distribution in the cavity for (11, 1) and (5, 2) modes modeled in COMSOL Multiphysics.

Conventionally, when TMDs monolayers are used as light emitters, the external resonator is tuned to the direct exciton transition\cite{ardizzone2019emerging, li2021experimental, sinev2021strong} (i.e., 780 nm/1.57 eV in the MoSe$_2$ case)\cite{tonndorf2013photoluminescence}. In our case, we use a bulk TMD layer simultaneously as a resonator and emitter. Therefore, the absorption of the emitted light is of great importance in our case. Even bulk TMDs have a strong absorbance near the excitonic resonance, while the luminescence is weak\cite{wilson1969transition,dong2015optical}. Thus,  to avoid maximum absorption, we should tune our resonators to the long wavelength tail of photoluminescence.

\begin{figure}[t!]
    \centering
    \includegraphics[width=\textwidth]{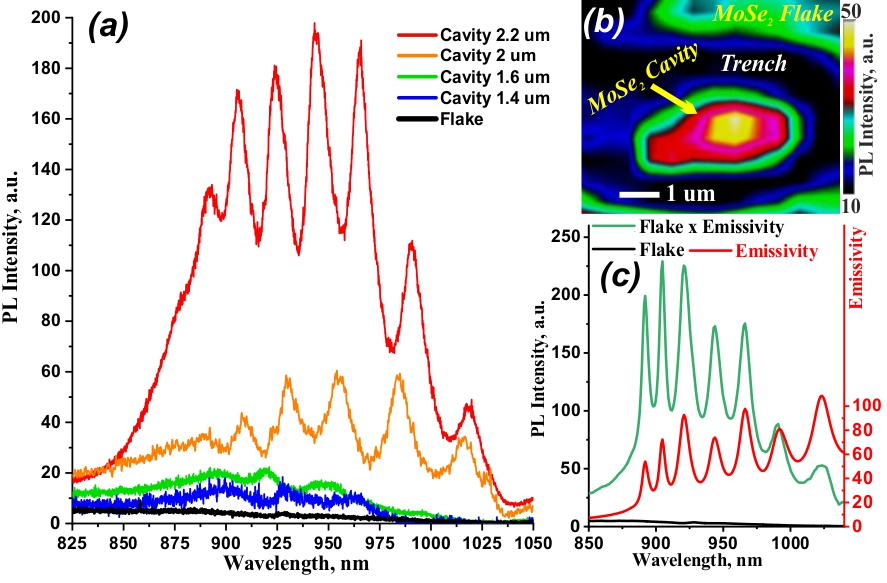}
    \caption{(a) PL spectra of cavities of various diameters. (b) PL mapping of one of the structures. (c) Calculated emissivity of 2.2 $\mu$m cavity (red curve), the flake spectrum (black curve), and the flake spectrum multiplied by the emissivity (green curve).}
    \label{diameters}
\end{figure}

Figure \ref{diameters} shows the $\mu$-PL data for cavities of various diameters. The spectra consist of a series of peaks specific for whispering gallery modes. Size variety of the cavities provides different positions of the maxima and enhancement factors. The 2.2 $\mu$m diameter cavity exhibits significantly enhanced PL with an enhancement factor of up to 100. Additionally, one can notice that the PL maxima of the cavities are shifted to the long-wavelength region comparing to the PL peak of the flake. Such a red-shift of the PL maxima can be explained by the dominant contribution of the cavity modes to the PL in the long-wavelength region (900-1050~nm) and suppression of the cavity radiation due to the significant MoSe$_2$ material losses in the short-wavelength region (800-900~nm). The PL maximum of the bare MoSe$_2$ appears in the highly absorptive wavelength range below 900~nm. Therefore, the cavity modes produce a small PL signal in the short-wavelength region where the absorptive losses are simultaneously resonantly enhanced. For this reason, the disks with smaller diameters of 1.4 and 1.6 $\mu$m having a high-Q resonances at the shorter wavelengths did not show such a strong enhancement\cite{borodin2021mechanical}. Moreover, we created structures of larger diameters (3, 5, and 10 $\mu$m). The PL data of these structures are presented in Supplementary Information in Figure S1. The 3 $\mu$m cavity demonstrates some features related to WGMs. 
However, these peaks are significantly weaker than in the 2.2 $\mu$m cavity case, and the flake-like PL peak dominates (see Figure 1S, red curve). With increasing cavity diameter, all WGM features disappear, and only the unmodified flake photoluminescence remains in spectra (see Figure S1, purple and blue curves). Such dependence of PL on cavity diameter can be explained by competition between the gain and material losses. On one hand, a larger cavity should provide higher enhancement due to the high-quality factor of modes with a higher azimuthal number. On the other hand, a longer light path in larger cavities relates to more intensive absorption by indirect bandgap material. Consequently, small cavities have low Q factor but low losses (1.4 and 1.6 $\mu$m), large cavities have high Q factor but high losses (3, 5, and 10 $\mu$m), and there is most likely some compromise with satisfying gain and losses (2, and 2.2 $\mu$m). Thus, we believe that the 2.2 $\mu$m diameter cavity is a compromise between miniature size and enhancement factor (~$\approx 100$).

The enhanced PL originates from the Purcell effect caused by the whispering gallery modes of the microdisk. The direct measurement of the Purcell factor through time-resolved PL is impeded due to the short (less than 1 ns) lifetime of excited states (see Figure 3S in Supplementary Information). To estimate an impact of the Purcell effect and validate our experimental results, we perform numerical calculations of emissivity of the microdisk in COMSOL Multiphysics\cite{dyakov2021photonic}. In order to calculate intensity of the emitted field, we utilize the reciprocity theorem. The reciprocity theorem dictates that the electric field intensity emitted by the current distribution is equal to the intensity necessary for excitation of the this current distribution. As a consequence, the ability of the cavity to absorb the incident energy matches with its ability to emit. Thus, to estimate the emissivity, it is enough to find the energy absorbed in the microdisk. Figure~\ref{diameters}(c, red curve) shows the emissivity spectrum of the microdisk with the diameter equal to 2.2 $\mu$m. The emissivity is calculated as an integral of the field intensity inside the microdisk. Multiplication of the emissivity and the flake PL spectrum gives the PL of the MoSe$_2$ microdisk. The positions of the peaks obtained in the calculations perfectly match with those are experimentally measured.

One can notice that conventionally WGM resonators do not emit vertically. However, the presence of defects might serve as a scatterer or antenna, and a finite aperture of the collecting objective provides simultaneous collection in some angle range. We can see such defects in our structures (e.g., non-uniform thickness in Fig. \ref{fabrication}d). In addition, we have noticed that the PL signals strongly depend on a collection spot. Figure 4b demonstrates the PL mapping of one of the structures. It can be seen that the PL intensity is not uniformly distributed, and there are hot spots. So, in our experiments, we chose the spots demonstrating the maximum PL intensity to record the PL spectra.

\begin{figure}[b!]
    \centering
    \includegraphics[width=0.9\textwidth]{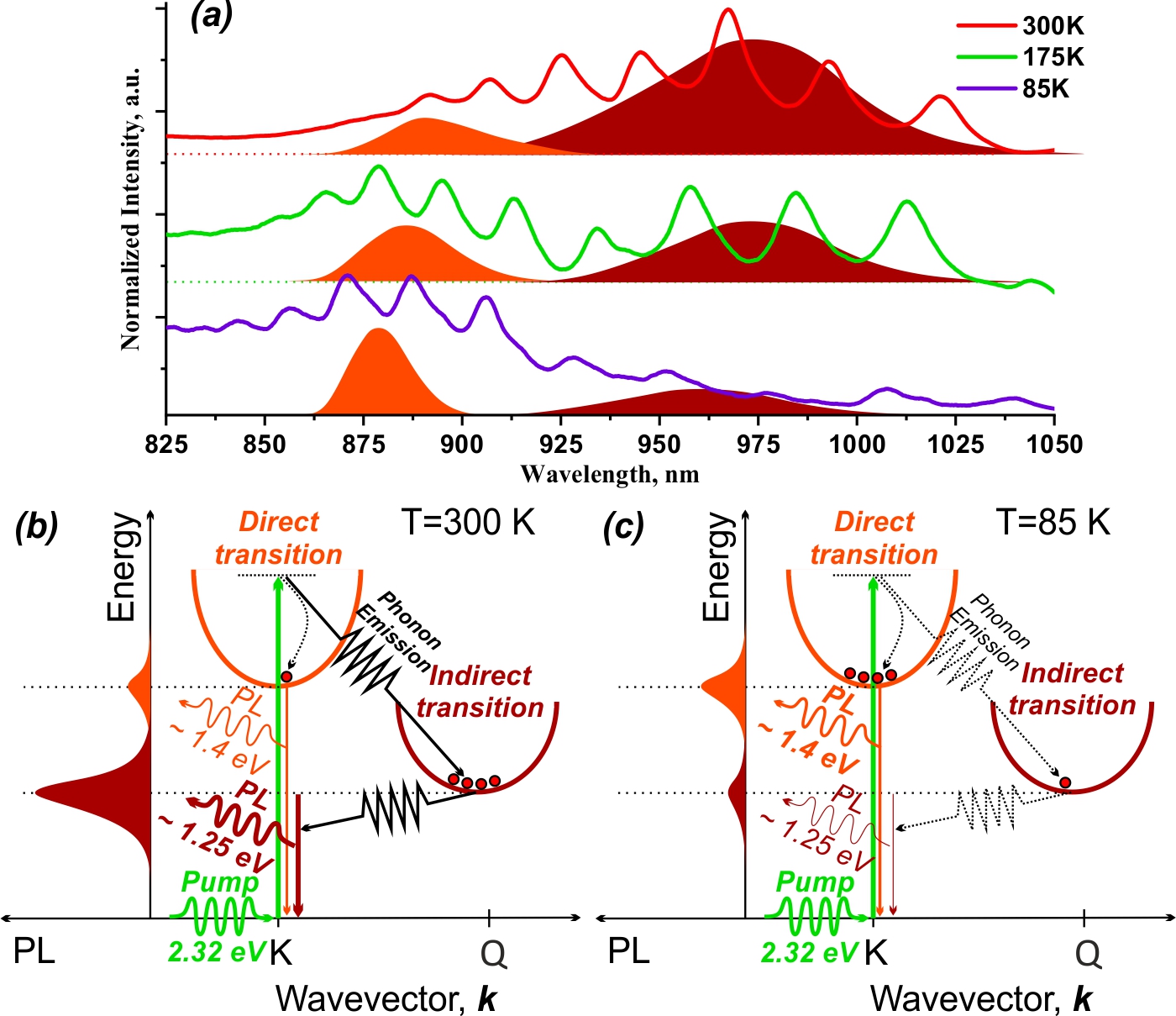}
    \caption{(a) Temperature dependence of photoluminescence. (b, c) Scheme of optical transitions at 300 K and 85 K, respectively. Orange and red contours are simple guides to the eye for direct and indirect transitions, respectively. Conventional fitting by Gaussian or Lorentzian curves is hampered due to the fact that PL spectra modified by WGM resonance, and PL of the thick flake is very weak.}
    \label{TPL}
\end{figure}

The nature of PL of such multilayer TMDCs structures is still ambiguous. However, many works point out the optical activity of both direct and indirect transitions in multilayered TMDs\cite{shubina2019excitonic,malic2018dark,smirnova2020temperature,brem2020phonon}. The common conclusion is that temperature plays a key role in the activation of an indirect transition. In our case, at room temperature, we have a broad PL spectrum modified by a WGM resonator; therefore, its analysis is nontrivial. For this reason, we investigated the temperature dependence of PL for our structures. The results of the investigation are shown in Figure \ref{TPL}.
As you can see from Figure \ref{TPL}a, with the decreasing temperature, the broad spectrum at 300 K (see red line) splits into two components at 175 K (see green line), and finally, at 85 K (see violet line), the long-wavelength peak quenches, and short-wavelength one becomes dominant in the spectrum. Based on such behavior of PL, we can assume that observed PL consists of two near peaks. The direct transition is about 1.4 eV, and the indirect transition is about 1.25 eV. The energy of excitation is much higher than both of these (i.e., excitation is non-resonant). Therefore, at room temperature, most of the excited electrons experience phonon-assisted relaxation to the lowest energy state. There, electrons commit radiative phonon-assisted recombination that provides the peak at 1.25 eV. Although, some electrons relax directly to a higher energy state that has a shorter lifetime (i.e., higher oscillator strength) and recombine directly without a phonon that provides the peak at 1.4 eV. This case is illustrated in Figure \ref{TPL}b that corresponds to the spectrum at 300 K (see Fig. \ref{TPL}a). With decreasing temperature, the PL peaks become narrow, and phonons start to freeze out that reduces the effectiveness of the phonon-assisted recombination channel. As a consequence, we can see the separation of direct and indirect peaks and the formation of a "trough" at 930 nm between them (see Fig. \ref{TPL}a, 175 K). At 85 K, phonons are almost frozen out, and the phonon-assisted channel is ineffective. Therefore, the indirect peak barely can be seen, and the direct channel becomes dominant. This case is illustrated in Figure \ref{TPL}c that corresponds to the spectrum at 85 K (see Fig. \ref{TPL}a). Although, as discussed earlier, the cavities enhance mainly the tails of the PL peaks. So, the actual energies of transitions are probably slightly higher than those we observe here.

\section{Conclusions}
To conclude, we investigated the optical properties of nanophotonic WGM disk resonators. The resonators were fabricated from bulk indirect bandgap MoSe$_2$ via resistless mechanical scanning probe lithography. The diameter of cavities varied from 1.4 to 10 $\mu$m, and the thickness was 70 nm. Micro-photoluminescence investigation revealed WGM-like PL spectra with various enhancement factors depending on the cavity diameter. It was shown that the optimal cavity diameter is 2.2 $\mu$m, which provides enhancement factor of ~$\approx 100$ compared to the pristine flake. Scattering experiments and modeling also revealed WGM spectra and confirmed data obtained by micro-PL. Moreover, we investigated the temperature dependence of a cavity PL. The results showed that PL has two components - the short-wavelength (~$\approx 885$ nm) and the long-wavelength ($\approx 990$ nm). With decreasing temperature, the long-wavelength component quenches, while the short-wavelength one becomes dominant.  Based on it, we assume that two transitions are simultaneously active in our structures PL - direct ($\approx 1.4$ eV) and indirect ($\approx 1.25$ eV). 

Thus, in this work, we demonstrated a novel approach to the fabrication of light-emitting nanophotonic devices based on bulk indirect bandgap TMDs and investigated the optical properties of several. We believe that this approach might be promising to create other light-emitting nanophotonic devices from bulk TMDs and reveal their fascinating properties.
\section{Methods}

\textbf{Sample preparation}.
 Thin-film MoSe$_2$ samples were fabricated by mechanical exfoliation with adhesive tape (blue tape, Nitto) from a commercial bulk crystal (obtained from HQ Graphene, Netherland) on top of the Au/Si substrate. 

\textbf{Cavities creation}.
The cavities were created by resistless frictional mechanical probe lithography. To perform the procedure, we used Ntegra Aura (NT-MDT) atomic force microscope using DCP (NT-MDT) probes with a curvature radius of 100 nm and a spring constant of 30-85 N/m. The multi-pass frictional approach was used to prevent defect formation. The force was about 10 uN, and number of passes amounted to 200 for each resonator. Detailed information on this approach and the cavity creation process is available in our previous works.\cite{borodin2021mechanical, borodin2021mechanical2}

\textbf{Micro-photoluminescence ($\mu$-PL) investigation}.
The optical properties of the structures were investigated by measuring the PL spectra. For these experiments, a multi-functional optical complex Horiba LabRAM HREvo UV-VIS-NIR-Open equipped with a confocal microscope was used. Spectra were obtained with a spectral resolution of ~3 cm$^{-1}$ using a 600 gr/mm grating. We used an Olympus MPLN100$\times$ objective lens (NA = 0.9) to obtain information from an area with a diameter of 1 $\mu$m. Apart from local measurements, PL mapping with spatial resolution of 0.5 $\mu$m was performed at the same setup using a motorized table. The measurements were performed with continuous-wave (cw) excitation using the 532 nm laser line of a Nd:YAG laser (Laser Quantum Torus). To prevent damage to the structures, the incident laser power was limited to 1 mW.

\textbf{Scattering experiments}.
The microdisks were illuminated with white polarised light (Ocean Optics HL-2000-HP in combination with linear polariser) at an incidence angle of 65 degrees with a low-aperture lens Mitutoyo Plan Apo NIR 10x 0.26 NA. The radiation scattered by the disks is collected using the Mitutoyo Plan Apo NIR 50x 0.65 NA which was analysed on a Horiba LabRAM HR 800 UV-VIS-NIR spectrometer. The numerical apertures of the lenses and the angle of incidence are chosen so that pump not pass through the collection channel.

\begin{acknowledgement}
There is no funding to report.

We thank Mikhail M. Glazov and Ivan V. Iorsh for fruitful discussions.
\end{acknowledgement}

\begin{suppinfo}

\end{suppinfo}

\bibliography{bibliography}

\end{document}


\newpage
Figure 1S demonstrates PL spectra of cavities with large diameters (3, 5, and 10 um).
\begin{figure}
    \centering
    \includegraphics[width=\textwidth]{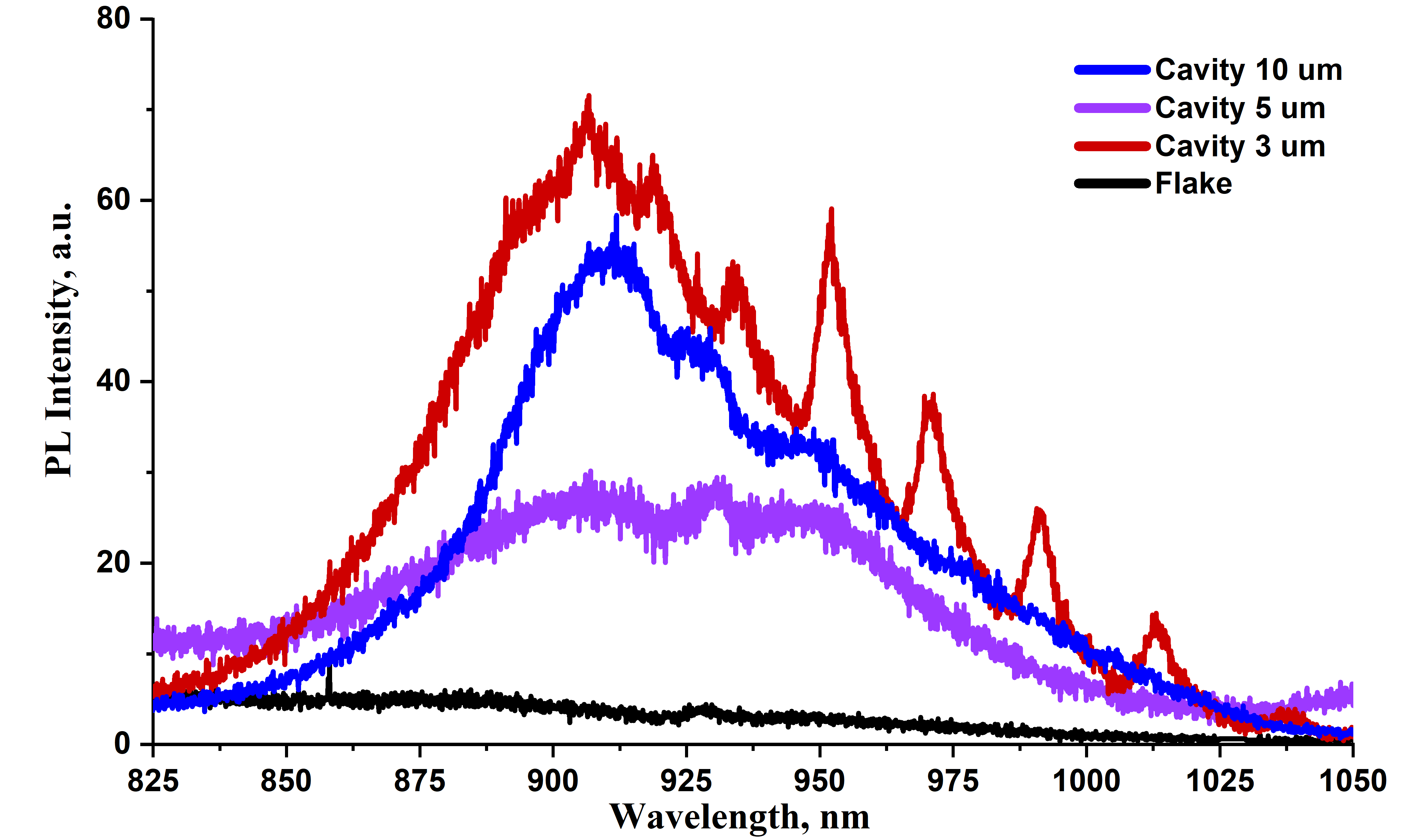}
    \caption{PL spectra of cavities of various diameters.}
    \label{PL}
\end{figure}
As one can see, the 3 um cavity demonstrates some features related to WGM resonance. However, these peaks are significantly weaker than in the 2.2 um cavity case, and the flake-like PL peak dominates (see Figure 1S, red curve). In the case of 5 and 10 um resonators, even weak WGM-like features disappear, and only unmodified flake photoluminescence remains in spectra (see Figure 1S, purple and blue curves).
\newpage
Figure 2S shows dependence of the maximum PL intensity on the cavity diameter.
\begin{figure}
    \centering
    \includegraphics[width=\textwidth]{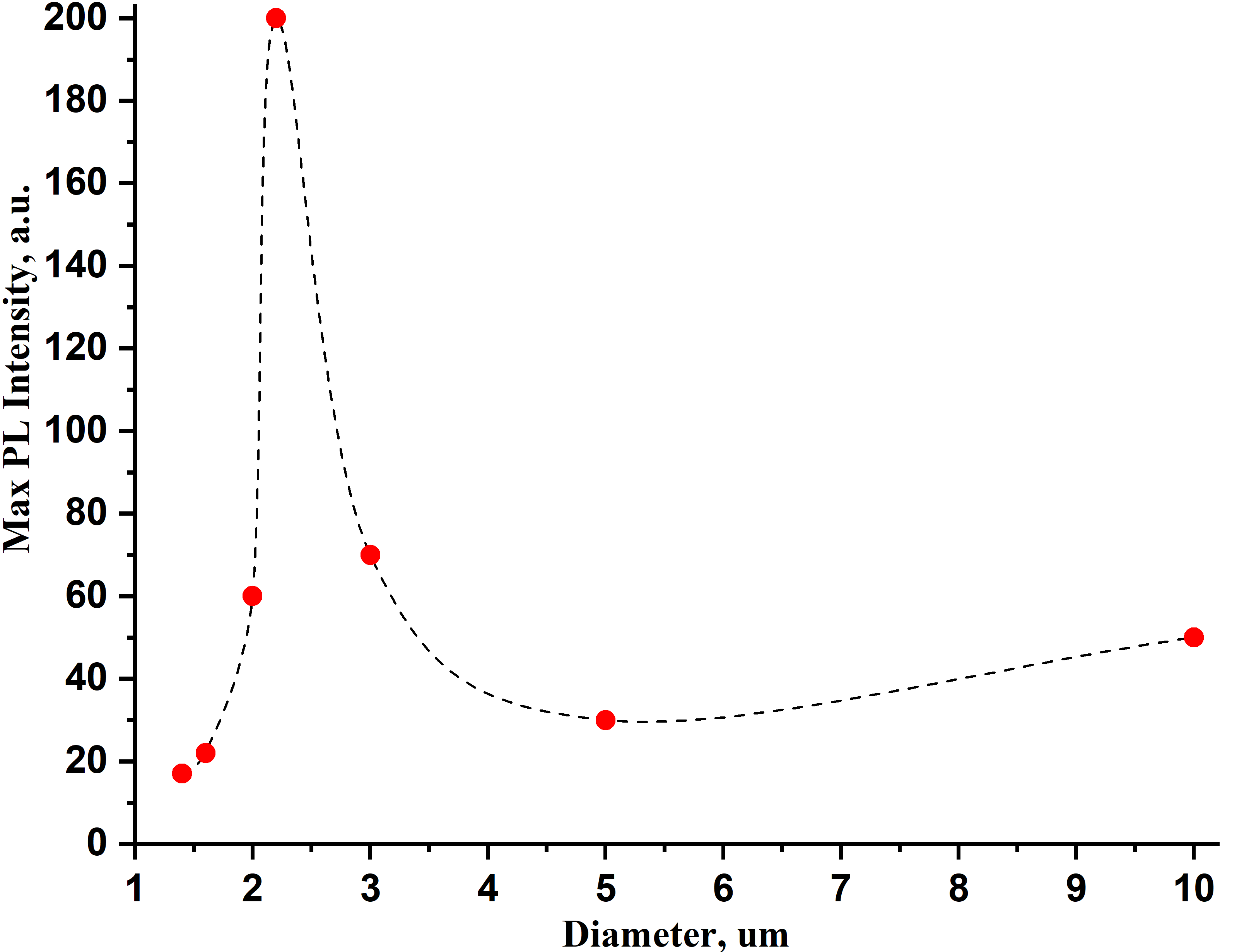}
    \caption{Dependence of the maximum PL intensity on the cavity diameter.}
    \label{PL}
\end{figure}

\newpage
Figure 3S demonstrates the results of the time-resolved PL study. As one can see, curves for the bulk flake and cavity are almost identical, and the curve of the laser itself (device-determined minimum curve) is very close to them. This means that the lifetime of excitement states is less than 1 ns, and our device does not have enough time resolution to detect it.

\begin{figure}
    \centering
    \includegraphics[width=\textwidth]{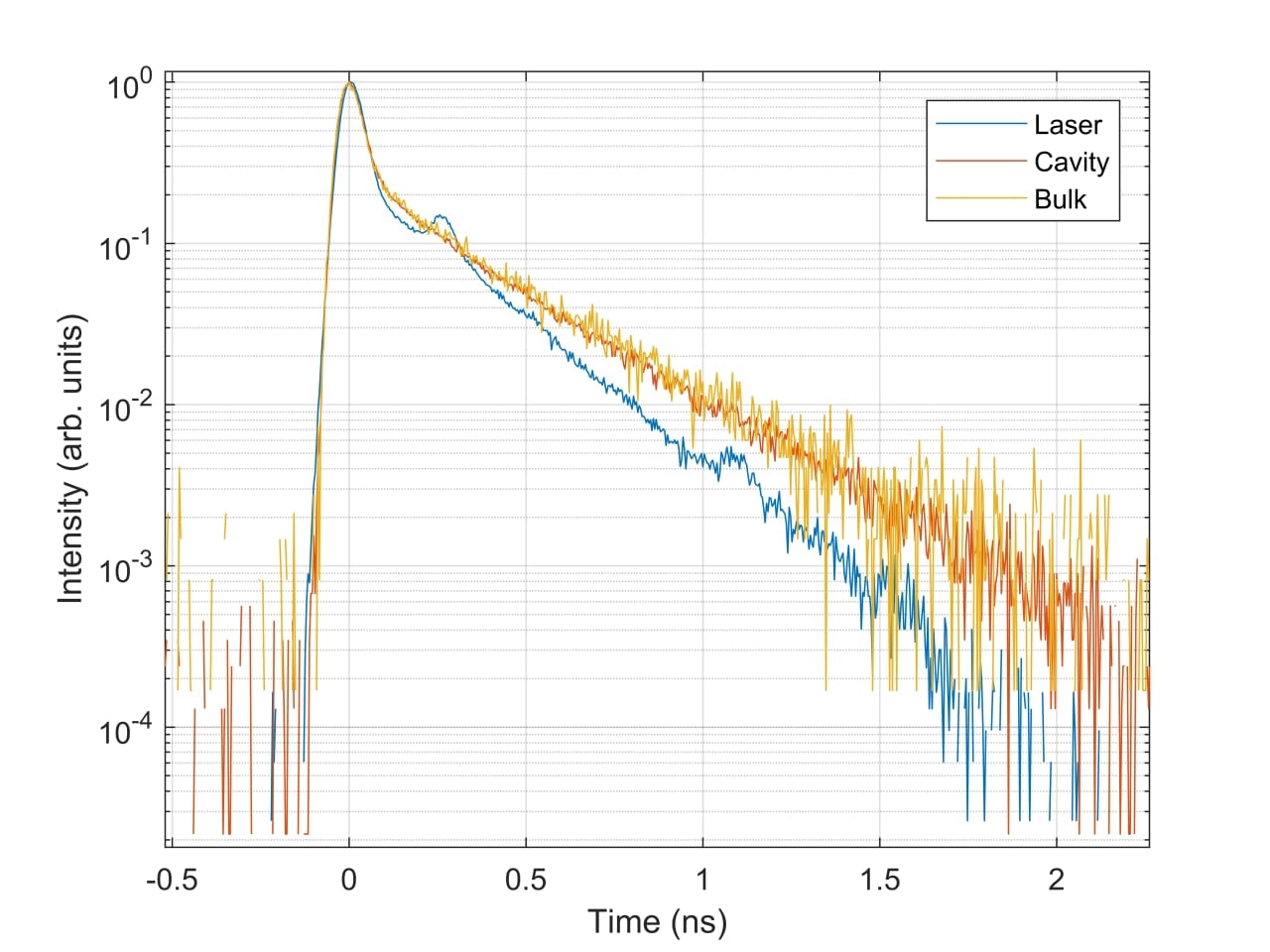}
    \caption{Time-resolved PL of the bulk flake (yellow curve) and cavity (red curve).}
    \label{PL}
\end{figure}